# Development of Hyperthermia Measurable Fiber Radiometric Thermometer for Monitoring Tissue Temperature during Thermotherapy

Paris L. Franz, Hui Wang*

Department of Chemical, Paper and Biomedical Engineering, Miami University, Oxford, OH, USA

**ABSTRACT:** Temperature monitoring is extremely important during thermotherapy. Fiber-optic temperature sensors are preferred because of their flexibility and immunity to electromagnetic interference. Although many types of fiber-optic sensors have been developed, it remains challenging for clinically adopting them. Here, we report a silica fiber-based radiometric thermometer using a low-cost extended InGaAs detector to detect black body radiation between 1.7µm to 2.4µm. For the first time, this silica fiber-based thermometer is capable of measuring temperature down to 35°C, making it suitable for seamless integration with current silica fiber catheters used in laser interstitial thermotherapy to monitor hyperthermia during a surgery. The feasibility, capability, and sensitivity of tracking tissue temperature variation were proved through *ex vivo* and *in vivo* tissue studies. The technology is promising for being translated into clinics after further improving the signal to noise ratio.

Temperature monitoring is essential for various applications, as such, a variety of temperature sensing methods have been developed[1]. Among them, fiber-optic temperature sensors have many unique features, such as flexibility, complete immunity to interference from radio frequency(RF) and microwave radiation, intrinsic reliability in harsh and corrosive environments, etc[2]. In general, fiber-optic temperature sensors can be categorized as structure-based, material-based, or radiation-based sensors. Fiber Bragg Grating (FBG) sensors are popular structure-based fiber-optic temperature sensors. FBGs are made by fabricating a fine volume grating in a fiber, and detect temperature related spectral shift[3,4]. Fluoroptic sensors are an example of material-based fiber-optic temperature sensors. The sensors are made by adding a fluorescent material on the tip of a fiber and detect temperature by measuring temperature induced fluorescence lifetime decay, spectral shift, or the intensity ratio from two different emission bands[5,6]. Pyrometer fiber sensors, a radiation based temperature sensor, can measure the black body radiation emitted from very hot surfaces (>300°C)[7,8]. They are usually used in extremely harsh environments which other sensors cannot access. For detecting temperatures lower than 100°C, pyrometer fiber sensors usually require special infrared fibers to transfer blackbody radiation (BBR) at midrange infrared wavelengths (MIR, ~3µm - ~8µm)[9–11].

Temperature monitoring is extremely important during thermotherapy in order to achieve effective and efficient treatment. Thermotherapy is a minimally invasive technique, which has been extensively adopted for clinical practices[12]. In thermotherapy, energy, such as RF[13], microwave[14], or laser[15,16], is delivered to diseased tissues to heat and then kill them without damaging surrounding normal tissues. The fundamental mechanisms of thermotherapy have been attributed to apoptosis and necrosis[17,18]. Apoptosis is a programmed cell death pathway, which occurs in a tissue temperature range between 40°C-45°C without inducing significant inflammatory activities. Necrosis occurs when tissue is heated to above 50°C. The disruption of the plasma membrane and the release of intracellular contents may lead to inflammatory and immunogenic responses. Except for vaporizing tissue[16], tissue temperature during thermotherapy is usually controlled between 40°C-80°C.

Fiber-optic temperature sensors are immune to interference from electromagnetic radiation and are preferably used for monitoring tissue temperature during thermotherapy[19]. However, most fiber-optic sensors are required to be in contact with tissues, except radiation-based sensors. In addition, each of them has its own limitations. FBG sensors are sensitive to strain and pressure induced by the motion of human body, such as respiratory movements, making clinical use challenging[20]. Fluoroptic sensors suffer artifacts due to self-heating when used during laser thermotherapy[21,22]. In principal, a radiation-based fiber-optic sensor is very attractive because it does not require contact with tissues. In particular, for laser interstitial thermotherapy (LITT)[23], the same fiber can be used for both delivering therapeutic laser energy to tissues and collecting the blackbody radiation emitted from the tissues to monitor temperature. This can make the system simple, compact, low-cost, and reliable. However, silica fiber, which is used to build the fiber catheter for LITT, cannot transmit BBR in MIW[24], the wavelength range typically used for monitoring the temperature below 100°. Therefore, a radiation-based fiber-optic sensor capable of measuring tissue temperature down to 40°C using a silica fiber as the probe is required for LITT.

In addition to fiber-optic temperature sensor, magnetic resonance imaging(MRI) thermometry has been recently translated into clinics[25,26]. MRI thermometry can non-invasively provide tissue temperature maps in 2D coregistered on MRI anatomy images, which enables sophisticated surgeries, such as treating epilepsy and glioblastoma through LITT[27–29]. In LITT, a therapeutic laser (10-12W) is delivered through a silica fiber to ablate targeted tissue under the guidance of MRI images and corresponding temperature maps[29]. Although, clinically, this strategy has shown very promising outcomes, the FDA recently issued letters warning that MRI thermometry may not be able to track tissue temperature variation in real-time due to long MRI acquisition time of 8 seconds, resulting in under-treatment or over-treatment[30–32]. The key issue is that during the silent 8 seconds, surgeons are blind to the temperature variation of the targeted tissue. If a sensor can update the targeted tissue temperature variation in real-time during two consecutive MRI temperature maps, the surgery will be more controllable. A radiation-based fiber-optic sensor is ideal for satisfying this requirement as no modification needed to the surgical fiber catheter side except adding BBR detection. In addition, surgeries under MRI guidance remain complex and costly, which may not be suitable for all cases[19]. For LITT, a fiber-optic temperature sensor, which can provide tissue temperature and be seamlessly integrated with a surgical fiber catheter, is desirable.

With aforementioned unmet clinical needs, in this paper, we aim to develop a silica fiber-based temperature sensor, called hyperthermia measurable fiber radiometric thermometer (hmFRT), which can measure temperature down to at least 35°C through BBR detection.

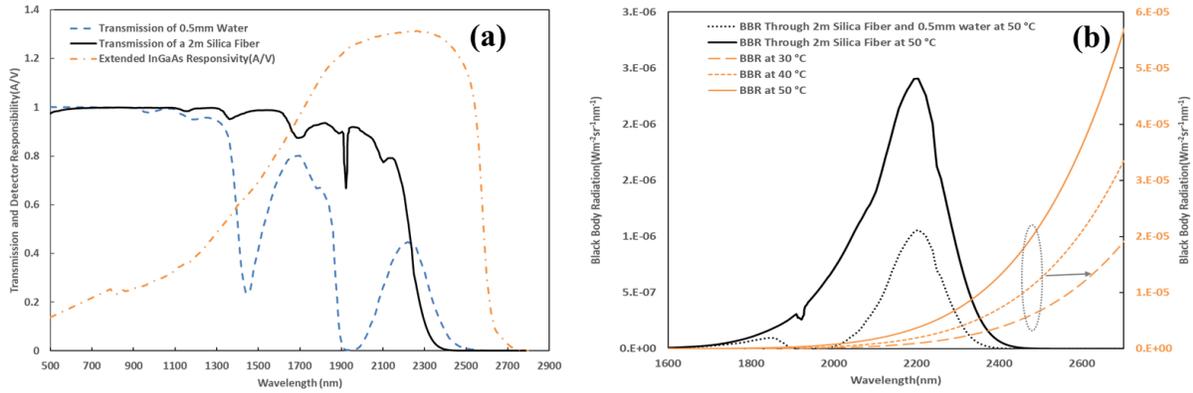

**Figure 1.** Optimal wavelength range for BBR detection with a silica fiber. (a) The spectral transmission of a 2m silica fiber, a 0.5mm water layer, and the spectral responsivity of extended InGaAs detector. (b) BBRs calculated using Plank's equation at 30ºC, 40ºC, and 50ºC and the BBR at 50ºC after passing through a 2m silica fiber and after passing through both a 2m silica fiber and a 0.5mm water layer.

To be able to measure the BBR emitted from a subject and transmitted through a silica fiber, we need to identify an optimum wavelength range for detecting the BBR. Figure 1(a) shows the spectral transmission of 2m silica multimode (MM) fiber, where the transmission is cut off at ~2.4µm[24]. It is possible that the targeted tissue will be covered with a thin layer of water, so we also plot the spectral transmission of a 0.5mm water layer in Fig.1(a), in which the cutoff wavelength is ~2.5µm[33]. Figure 1(b) plots the BBR calculated with Planck's law at three different temperatures, 30ºC, 40ºC and 50ºC. Based on the spectral transmission in Fig.1(a), we calculated the BBR at 50ºC after being conducted through a 2m silica fiber. As shown in Fig.(b), the maximum transmission is between ~1.7µm – ~2.4µm peaked at ~2µm. If a 0.5mm water layer is added, the maximum transmission resides in two bands at ~1.8µm and ~2.15µm. For either of these cases, the optimum detection wavelength range is between ~1.7µm – ~2.4µm. Among all available photodetectors working at room temperature, the extended InGaAs detector has the maximum responsivity in this wavelength range as shown in Fig.1(a) compared with other infrared photodetectors, such as PbS or InAsSb[34]. Based on this analysis, we developed an hmFRT to detect BBR between ~1.7µm – ~2.4µm though a silica fiber with a low-cost extended InGaAs detector, and then verified its feasibility for thermotherapy applications.

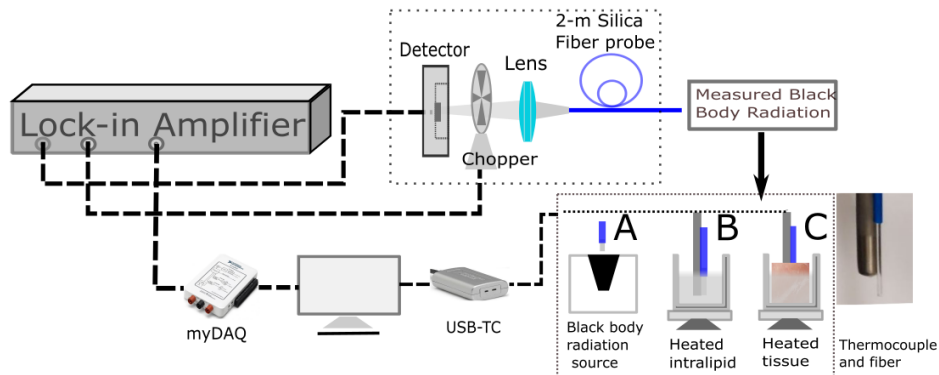

**Figure 2.** Experimental layout of the hyperthermia measurable fiber radiometric thermometer. Black body radiation was measured from three different setups as shown in the dash line box. A) Measured from a standard black body radiation source. B) Measured from heated 1% intralipid. C) Measured from a piece of heated tissue. For (B) and (C), the fiber probe was bound with a thermocouple.

Figure 2 shows a schematic diagram of the hmFRT system. The BBR was measured through a 2-meter silica fiber probe under three different cases, A, B, and C, as shown in the dashed-line box. For case A, a standard BBR source (Optris BR400) was used to calibrate the hmFRT at different temperatures in order to obtain a calibration curve. For cases B and C, the fiber probe was bound with a type K thermocouple probe (MacMaster) as shown in the photo in Fig.2. The bound fiber probe and the thermocouple were inserted into 1% heated intralipid (case B) to simultaneously measure the BBR signal and the temperature of the intralipid solution during a heating process. The 1% intralipid was used to simulate a tissue environment. To verify that the hmFRT can work in tissue, we further tested the hmFRT by burying the bound fiber probe and the thermocouple into a piece of chicken breast. The tissue was treated with two heating and cooling cycles to test if the FRT can accurately track the temperature variation during the process. The BBR was collected and transmitted through a 2m MM silica fiber with a core diameter of 600µm (Thorlabs, FP600ERT). Then the BBR was focused through a mechanical chopper (Stanford Research Systems, SR540), which modulated the BBR to 1.5kHz. A low-cost extended InGaAs detector (Thorlabs, DET10D2) with a spectral responsivity from 0.9µm to 2.6µm was used to convert the BBR to voltage signal. The detector worked in photovoltaic mode to keep the dark current at a minimum. The modulated BBR signal was amplified, demodulated, and filtered by a lock-in amplifier (Stanford Research Systems, SR530). The reference signal for demodulation was obtained from the optical chopper and the time constant of the lock-in amplifier was set at 1 second. After demodulation, the signal was digitalized with a USB data acquisition card (MyDAQ, National Instruments). Simultaneously, the signal from the thermocouple was sampled as a reference of the true temperature for cases B and C by a USB-TC (National Instruments). We also calibrated the thermocouple with the resistance temperature detector associated with the standard BBR source. The software for data acquisition was developed with Labview (National Instruments).

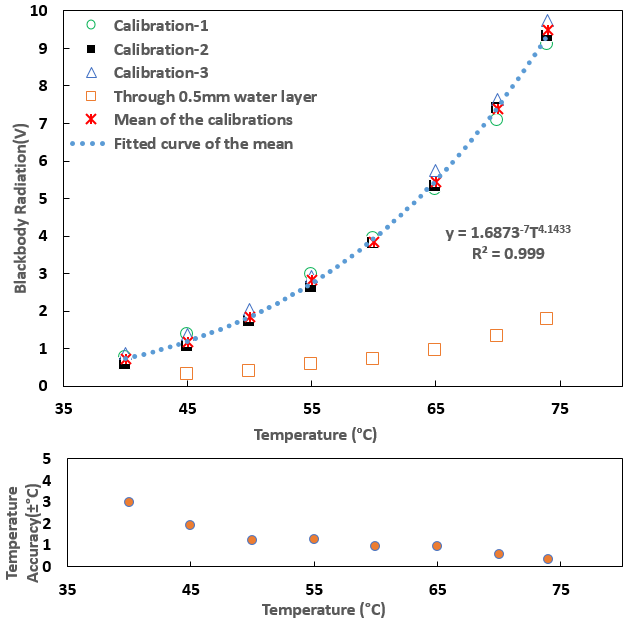

Figure 3. The hmFRT was calibrated three times with the standard black body radiation source. (a) The mean values of three measurements were fitted with a power function. In addition, a 0.5mm water layer was placed ahead of the fiber probe and the measured signals (boxes) are also plotted in the figure. (b) The standard deviation of the BBR signal was converted into the temperature accuracy using the calibration equation.

The hmFRT was first calibrated with a standard BBR source to quantify the relation between the true temperature and the measured BBR signal. We put the fiber probe in front of the BBR source at three different distances (~5mm, ~8 mm, ~12mm) and recorded the BBR signals at seven temperatures from 40°C to 74°C. The measured data are shown in Fig.2(a) as calibration 1-3. We see the BBR signals are insensitive to distance, which is consistent with previous theoretical studies that fiber radiometric thermometer is insensitive to the distance between a fiber end to a thermal surface[35]. The mean values of the three measurements can be fitted by a power function very well as indicated by the correlation coefficient, R. Based on the fitted equation, the calibration equation to convert the BBR signal to the corresponding temperature is:

$$T(°C) = C \cdot 43.118 \cdot B^{0.2412} \qquad [1]$$

where B is the measured BBR signal in voltage (V) and C is a correction factor, where C=1 for case A when using a standard BBR source. A correction factor is needed as the calibration with the standard BBR source is conducted in air, while in real applications, this may not be the case. For example, the fiber probe may be in direct contact with tissue or in a cooling solution, in which Fresnel reflection from the interface between the silica and tissue and the emissivity of the tissue will be different from those during the calibration.

As shown in Fig.1, the water absorption at 1.7μm and 2.2μm is much lower than that in long infrared range (3μm -10μm), the spectral range typically used for thermal imaging. It is possible to sense the BBR signal with our method even when there is a thin layer of water, which may be used in surgeries as a cooling solution. Therefore, we put a water filled cuvette with a thickness of 0.5mm between the fiber probe and the BBR source. We were able to detect the BBR signal down to 45 °C as shown in Fig.2(a).

The temperature measurement accuracies of the hmFRT can be calculated based on the standard deviations (std) of the measured BBR signals at different temperatures. With Eq.1, we are able to convert the BBR signal to a corresponding temperature. Assuming the conversion between the BBR signal and the temperature is linear, which is true in a small range around the measured BBR signal, we can plot the measurement accuracies of the FRT at different temperatures, shown in Fig.2(b). The accuracy is much better at higher temperatures (±0.4°C at 74°C), than at lower temperatures (±3°C at 40°C), because the BBR signal is significantly increased with the increase of the temperature resulting in a better signal to noise ratio.

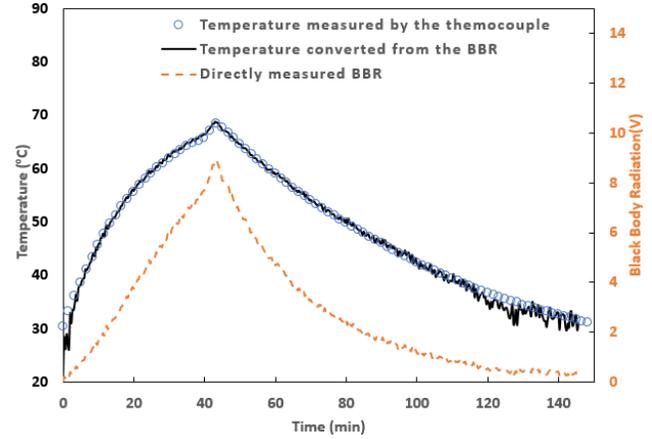

Figure 4. The BBR signal measured with the hmFRT and the temperature simultaneously measured with the thermocouple within 1% heated intralipid. The measured BBR signal is converted to temperature with the calibration equation [1] obtained in Fig.2.

We were able to test the performance of the FRT and obtain the calibration equation using the standard BBR source. However, during thermotherapy, the fiber may be inserted into tissue. It is important to calibrate and test the hmFRT within a tissue phantom. We bound the fiber probe end with the thermocouple and immersed them in a 1% intralipid solution, in order to acquire both the BBR signal and the true temperature of the intralipid solution simultaneously. The intralipid solution was heated with a temperature-controlled oven (Corning PC-4200). Figure 4 plots the variation of the BBR signal and the measured temperature from the thermocouple through one heating and cooling cycle from the room temperature. Although we sampled data every second, in Fig.4 for clarity, the BBR signal was plotted every 20s, while the temperature was plotted every 100s. Then using Eq.1, we can convert the BBR signal to the corresponding temperature by multiplying by a correction factor C = 0.937. The correction factor was calculated by fitting the converted temperatures using Eq.1 to the measured temperatures using least squares minimization. From Fig.4, with the correction factor, the BBR can track the temperature change well from ~ 35 °C to ~68°C. The maximum measured temperature by FRT is limited by the maximum voltage level that can be measured by the DAQ card (10V).

After calibrating the hmFRT within a phantom, the intralipid solution, we conducted an *ex vivo* tissue study where the fiber was buried in a chicken breast. As shown in Fig.2 case C, the fiber probe end and the thermocouple were buried into the tissue together, so we could verify that the FRT can track the temperature variation of the tissue. Figure 5 plots the variations of the BBR signal and the measured temperature of the tissue during two heating and cooling cycles, one slow cycle and one fast cycle. We then converted the BBR signal into the real temperature using Eq.1 and the correction factor, C = 0.937, obtained in the phantom study. The system can track the variation of the tissue temperature closely from ~ 35°C to ~70°C similar to the results shown in Fig.3. The BBR signal was saturated at ~ 71°C, shown as a flat top in Fig.5. During the sharp temperature increase from the 38[th] minute to the 41[th] minute, the temperature increases by 30 ºC. The hmFRT is capable of tracking temperature change at 10ºC/minute. The inset photo in Fig.5 shows the tissue appearance after the two cycles of heating and cooling. In this study, we did not introduce an ablation laser beam to exactly simulate LITT. However, this study simulates the scenarios of tissue ablation through RF ablation or microwave ablation.

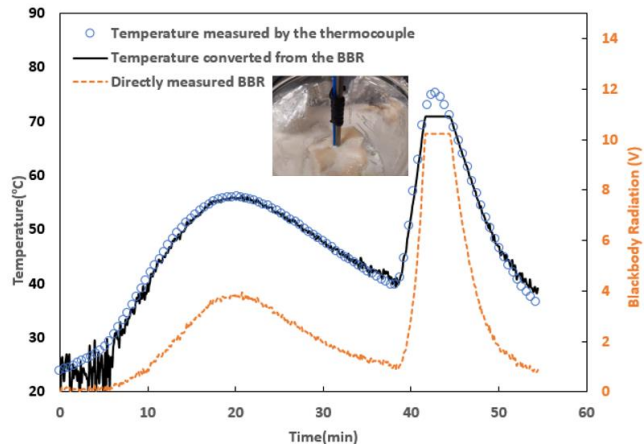

Figure 5. The BBR measured with the hmFRT and the temperature simultaneously measured with the thermocouple by inserting both into chicken muscle to simulate interstitial thermal therapy. The measured BBR is converted to temperature with the calibration equation [1].

To further demonstrate the sensitivity of the hmFRT, we tried to measure the core body temperature of a volunteer. The fiber probe was placed into the mouth cavity of the volunteer and the BBR signal was recorded every 0.5s. Normally, human core body temperature is around 37°C. As shown in the gray box of Fig.6, the hmFRT can track the core body temperature well. The variation during the measurement can be attributed to the inherent noise induced by the electronic devices, including the detector, the lock-in amplifier and the DAQ, which is more significant at a lower temperature as shown in Fig.2(b).

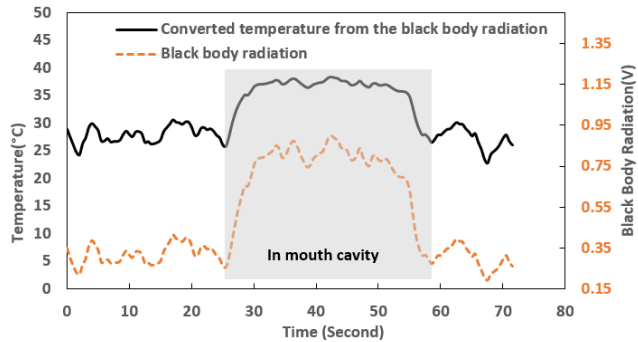

Figure 6. Black body radiation measured with the hmFRT by inserting the fiber probe into a volunteer's mouth cavity. The measured blackbody radiation is converted to temperature with the calibration equation obtained in Fig.2.

We have demonstrated that the hmFRT can detect BBR emitted from a tissue at temperatures down to 35ºC using a 2m silica fiber as the transmission media. After calibration, the hmFRT can accurately track tissue temperature change at a rate of 10ºC/minute. Such sensitivity and tracking ability show that hmFRT is promising for detecting two important hyperthermia temperature ranges, 40ºC -45ºC and 50ºC-80ºC, during thermotherapy. This feasibility has not been demonstrated before but open a new way to monitor temperature in real-time during thermotherapy. Compared with the popularly used FBG and fluoroptic sensors, this radiation based hmFRT not only inherits the desirable features of fiber-optic sensors but has several unique advantages. It can be seamlessly integrated with current LIIT without adding any complexities to the surgical fiber catheter. The same fiber can be used for both treatment and temperature sensing by simply adding a lock-in detection system at the detection end. In addition, the hmFRT can measure temperature without direct contact with the tissue, which could be very important for applications where the therapeutic fiber is encapsulated in a protection sheath[29]. Furthermore, temperature sensing can be combined with endoscopic optical imaging, such as optical coherence tomography, by using a dual cladding or a dual core fiber to acquire both images and a tissue surficial temperature map[36].

Currently, the major limitation for the hmFRT demonstrated here is that the signal to noise ratio (SNR) may not be sufficient for real-time sensing and high accuracy measurements at lower temperatures, such as 40ºC. Usually, during thermotherapy, the accuracy of temperature measurement should be between 1-2 ºC, while currently hmFRT at 40ºC is around ±3ºC due to very low BBR signal at low hyperthermia temperatures. A 10dB SNR improvement may be required to increase the accuracy. For many clinical applications, such as in conjunction with MRI thermometry, it is important to be able to update temperature at every 0.1s, which requires another 10dB SNR improvement based on current 1s time constant set in the lock-in amplifier. Technically, a 20dB-30dB SNR improvement is possible. In our current setup, the BBR is only modulated at 1.5kHz. Therefore, the SNR suffers flicker noise. By improving the modulation frequency to above ~100kHz, we can expect ~10dB SNR improvement[37]. A low-cost extended InGaAs detector at room temperature was used here. The SNR could be further improved by using a cooled detector[34]. In addition, all optics used here were not optimized for the wavelength range between 1.7µm- 2.4µm. The optical transmission and coupling efficiency of detection can be further improved if optimized components used. Therefore, it is feasible to build a hmFRT system to sense temperature down to 40°C at a high temporal resolution by optimizing the system.

In this paper, we show that the optimal wavelength range for detecting BBR using a silica fiber as transmission media is between 1.7µm- 2.2µm. Based on this analysis, we have built an hmFRT system. Though *ex vivo* and *in vivo* tissue studies, we demonstrated that hmFRT can measure tissue temperature down to 35ºC and is a promising technology for monitoring hyperthermia during thermotherapy. This technology enables the possibility of using a single fiber for both treatment and temperature sensing during LITT. The hmFRT could also be a solution to the challenges faced by LITT for treating epilepsy and glioblastoma due to the slow updating rate of MRI thermometry.

## AUTHOR INFORMATION

**Corresponding Author**
*E-mail: hui.wang@miamioh.edu
**Present Addresses**
Paris Franz is currently with the Department of Applied Physics, Stanford University, Stanford, CA, USA

**Notes**
The authors declare the following competing financial interest: Miami University has filed a patent including the research findings disclosed in this manuscript.